\documentclass[prb,aps,floats,showpacs,amsmath,amssymb,preprint]{revtex4}
\usepackage[usenames]{color}
\usepackage{bm,graphicx,epsf}
\usepackage{bm,graphicx}
\usepackage{color}
\newcommand{\be}{\begin{equation}}
\newcommand{\ee}{\end{equation}}   
\newcommand{\bea}{\begin{eqnarray}}
\newcommand{\eea}{\end{eqnarray}}
\newcommand{\phrl}[1]{Phys.~Rev.~Lett. {\bf #1}}
\newcommand{\phrb}[1]{Phys.~Rev.~B {\bf #1}}

\newcommand{\jpsj}[1]{J.~Phys.~Soc.~Jpn.{\bf #1}}
\newcommand{\jpcm}[1]{J.~Phys.:~Condens.~Matter}
\newcommand{\bib}{\bibitem}
\newcommand{\lb}{\left[}
\newcommand{\rb}{\right]}
\newcommand{\lp}{\left(}
\newcommand{\rp}{\right)}

\renewcommand{\k}{{\bf k}}

\begin{document}

\title{Superconductivity in  non-centrosymmetric LaNiC$_2$}
\author{Soumya P. Mukherjee and Stephanie H. Curnoe}
\affiliation{Department of Physics and Physical Oceanography, Memorial University of Newfoundland,\\ St. John's, Newfoundland \& Labrador A1B 3X7, Canada }

\date{\today}


\begin{abstract}
We present a discussion of superconductivity in non-centrosymmetric superconductors with spin-orbit coupling.  A general expression for the 
quasiparticle excitation energy is derived.  The superconducting states 
for the point group $C_{2v}$ are classified by symmetry in the limit of weak spin-orbit coupling.  
A non-unitary triplet pairing gap function
can account for observations of broken time-reversal
symmetry and nodes in the superconducting state of LaNiC$_2$; however such a gap function also has a gapless branch and requires vanishing spin-orbit coupling.

\end{abstract}

\maketitle

\section{Introduction}

There are many non-centrosymmetric (NCS) superconductors, including CePt$_3$Si,\cite{Bauer} Cd$_2$Re$_2$O$_7$,\cite{Hanawa2001,Castallan2002} Mg$_{10}$Ir$_{19}$B$_{16}$,\cite{Klimczuk}  Li$_2$Pd$_3$B,\cite{Togano} Li$_2$Pt$_3$B,\cite{Badica}  UIr,\cite{Akazawa} LaNiC$_2$, \cite{Lee} and
BiPd.\cite{Joshi} 
The main consequence of broken inversion symmetry in superconductors is that it admits the possibility of mixed spin singlet and spin triplet pairing. However, the determination of the symmetry of the superconducting state follows the same principles as for centrosymmetric superconductors, namely the identification of the order parameter corresponding to an irreducible representation of the symmetry group of the crystal. The symmetry group of the normal state consists of the (magnetic) space group, $U(1)$ gauge symmetry, and $SU(2)$ spin rotation symmetry if spin-orbit coupling is sufficiently weak. Often,
the magnetic space group is just the 
product of the ordinary space group with time reversal symmetry (TRS); however in some cases 
(such as CePt$_3$Si) the magnetic space group is more complicated because of the existence of magnetic order above the superconducting transition. Here we are concerned with the former type, among them LaNiC$_2$.

The various phenomena associated with broken gauge symmetry mark the onset of superconductivity.  If gauge symmetry is the only symmetry that is broken at
the transition 
then superconductivity is ``conventional", otherwise it is ``unconventional".  In unconventional superconductors broken gauge symmetry may be accompanied by a lower point group symmetry, broken TRS, or even broken translation symmetry. There are few examples of superconductors with broken TRS.  The most famous is Sr$_2$RuO$_4$,\cite{Luke} and it has also been proposed for PrOs$_4$Sb$_{12}$.\cite{Aoki} A recent $\mu$SR experiment by Hillier {\em et al.} \cite{Hillier} showed that on entering the superconducting state at $T_c=2.7K$, LaNiC$_2$ also simultaneously breaks TRS.

Various experiments performed on LaNiC$_2$  have led to different proposals for the symmetry of the superconducting gap function based on the presence of 
nodes. 
Early measurements of specific heat \cite{Pecharsky} and NQR 1/T$_1$ relaxation time \cite{Iwamoto} suggested conventional BCS behaviour. 
Another early experiment on specific heat found evidence for point nodes.
\cite{Lee} 
Finally, a 
recent penetration depth measurement \cite{Bonalde} found a power law dependence on temperature, indicative of nodes, most likely line nodes. 
Based on these experimental findings, we will explore the possible symmetry groups of
the superconducting order parameter, especially those with broken TRS. 


The outline for this article is as follows.
We begin by examining band electrons in a lattice without inversion
symmetry with spin-orbit coupling.  Using a spin helicity basis, we derive a general
expression for the quasi-particle excitation energy.
In order to classify superconducting states in LaNiC$_2$, 
two approaches are considered in the limit of weak spin-orbit coupling.\cite{Quintanilla,Hillier}  The type (line vs.\ point) and location of    
 nodes are calculated and compared with experiment.

\section{Superconductivity in non-centrosymmetric crystals}
The normal state Hamiltonian for 
band electrons in a lattice without inversion symmetry is 
\be
H_0 = \sum_{\k, s}\xi_\k c_{\k s}^\dagger c_{\k s} + \sum_{\k,s,s'}\lp \bm{g}_\k \cdot \bm{\sigma}\rp_{ss'}c_{\k s}^\dagger c_{\k s'} \, ,
\label{H_normal}
\ee
where electrons with momentum $\k$ and spin $s\, (=\uparrow \text{or} \downarrow)$ are created (annihilated) by the operators $c_{\k s}^\dagger$ ($c_{\k s}$), $\xi_\k$ is the band energy measured from the Fermi energy $\epsilon_F$ and $\bm{\sigma}$ are the Pauli matrices. 
The second term in the Hamiltonian (\ref{H_normal}) 
is spin-orbit coupling (SOC).   
The form of the SOC 
is governed by the symmetry of the underlying point group.  
 In a non-centrosymmetric crystal it is common to assume
that $\bm{g}_{-\k} = -\bm{g}_\k$. 
Because of broken parity {\em and} SOC, the 
spin degeneracy of the band is lifted. 
By diagonalizing $H_0$, one finds two non-degenerate
bands with energies $\xi_{\k\lambda} = \xi_k +\lambda \vert \bm{g}_\k \vert$ where $\lambda = \pm 1$ is ``helicity" of the bands. 
Therefore in the diagonalized basis $H_0$ (\ref{H_normal}) becomes $H_0 = \sum_{\k , \lambda} \xi_{\k\lambda }\tilde{c}_{\k \lambda}^\dagger \tilde{c}_{\k \lambda} $, 
where $\tilde{c}_{\k \lambda}^\dagger$ and $\tilde{c}_{\k \lambda}$ are 
the electron creation and annihilation operators
for the band with helicity $\lambda$ and momentum $\k$. The unitary
transformation from the spin basis to the helicity basis is\cite{Mineev, Samokhin}
\bea
\nonumber
&& c_{\k \uparrow} =\frac{1}{\sqrt{2 |\bm{g}_\k|}} \lb \sqrt{|\bm{g}_\k|+g_{\k z}} \tilde{c}_{\k_+} + \sqrt{|\bm{g}_\k|-g_{\k z}} \tilde{c}_{\k_-}\rb \\ \nonumber
&& c_{\k \downarrow} = \frac{e^{i\phi_k}}{\sqrt{2 |\bm{g}_\k|}} \lb \sqrt{|\bm{g}_\k|-g_{\k z}} \tilde{c}_{\k_+} - \sqrt{|\bm{g}_\k|+g_{\k z}} \tilde{c}_{\k_-}\rb
\eea
where $e^{i\phi_{\k}} = \frac{g_{\k x} + i g_{\k y}}{|\bm{g}_{\k}|}$.

To describe superconductivity, we add to $H_0$ (\ref{H_normal}) a 
superconducting pairing term $H_1$,
\bea
\nonumber
&& H=H_0+H_1\\ 
&& H_1= \frac{1}{2}\sum_{k,s,s'} \lb \Delta_{ss'}(\k) c_{\k s}^\dagger c_{-\k s'}^\dagger - \Delta^{\ast}_{ss'}(-\k) c_{-\k s}c_{\k s'}\rb
\eea
where $\Delta(\k)$ is the gap function, a $2\times 2$ matrix in 
spin space of the form
\begin{equation}
\Delta(\k) = i\psi(\k)  \sigma_y, 
\,\,\,\,  \psi(\k) = \psi(-\k)
\end{equation}
for singlet spin pairing, or
\begin{equation}
\Delta(\k) = i{\bf d(\k)}\cdot \bm{\sigma}  \sigma_y, 
\,\,\,\, {\bf d}(\k) = -{\bf  d}(-\k)
\end{equation}
for triplet pairing. 
$\Delta(\k)$ is associated with one of the irreducible representations 
of the point group. 
In a NCS superconductor, the gap function can be a mixture of even and odd representations, 
\be
\Delta(\k)=\lb \psi(k) + {\bf d(\k)}\cdot \bm{\sigma}\rb i \sigma_y.
\label{delta}
\ee
If $\Delta\Delta^{\dagger}$ is proportional to the unit matrix then superconductivity is ``unitary", otherwise it is ``non-unitary".  
In a centrosymmetric superconductor, non-unitarity arises in the triplet
channel from broken TRS, which splits the quasi-particle
energy
degeneracy.
In a NCS superconductor, the gap function (\ref{delta}) is non-unitary due to 
mixed parity. 

In the helicity basis, $H$ takes the form
\be
H=\sum_{\k,\lambda} \lb \xi_{\k\lambda }\tilde{c}_{\k \lambda}^\dagger \tilde{c}_{\k \lambda}+\Delta_{\lambda}(\k) \tilde{c}_{\k \lambda}^\dagger \tilde{c}_{-\k \lambda}^\dagger-\Delta^*_{\lambda}(-\k)\tilde{c}_{-\k \lambda} \tilde{c}_{\k \lambda}\rb 
+ \sum_{\k , \lambda \lambda'} \lp \Delta_{\lambda \lambda'}(\k)  \tilde{c}_{\k \lambda}^\dagger \tilde{c}_{-\k \lambda'}^\dagger -\Delta^*_{\lambda \lambda'}(-\k) \tilde{c}_{-\k \lambda} \tilde{c}_{\k \lambda'}\rp,
\ee
where 
\bea
\label{del_l}
&& \Delta_{\lambda}(\k)=- \lambda e^{-i\phi}\lb \psi(k)+\lambda ({\bf d(\k)}\cdot \hat{\bm{g}}_\k)\rb \\ 
&& \Delta_{\lambda \lambda'}(\k)= \frac{\lambda e^{-i\phi}}{\sqrt{g^2_{\k x}+g^2_{\k y}}} [\{\hat{\bm{g}}_\k \times ({\bf d(\k)} \times {\bm{g}}_\k)\} + i \lambda' ({\bf d(\k)} \times {\bm{g}}_\k)]_z 
\label{del_ll'}
\eea
and 
$\hat{\bm{g}}_\k$ is the  unit vector along ${\bm{g}}_\k$. 
Eq.~(\ref{del_ll'}) clearly indicates that for ${\bf d}(\k)$ parallel to ${\bm{g}}_\k$ 
the interband pairing term completely disappears. This results in less condensation energy for the superconducting state therefore stabilizing 
${\bf d}(\k) \parallel {\bf g}_{\k}$.\cite{Sigrist}  

Neglecting interband pairing, we find 
the quasiparticle excitation energy,
\be
E_{k \lambda}= [ \xi^2_{k \lambda}+ |\psi(k)|^2 + ({\bf d(\k)}. \hat{\bm{g}}_\k)({\bf d^*(\k)}.\hat{\bm{g}}_\k) + \lambda (\psi(k) {\bf d^*(\k)} + \psi^*(k) {\bf d(\k)}).\hat{\bm{g}}_\k)]^{\frac{1}{2}},
\label{energy_NCS}
\ee
which is non-degenerate due to the lifting of band degeneracy in the normal state {\em and} a  mixed parity superconducting gap function.
When TRS is preserved (${\bf d}$ and $\psi$ 
are real),  we have
\be
E_{k \lambda}= \lb \xi^2_{k \lambda}+ \lp \psi(k) + \lambda ({\bf d(\k)}. \hat{\bm{g}}_\k) \rp^2 \rb^{\frac{1}{2}},
\ee
which is a well-known result for a general NCS superconductor.
However, in
LaNiC$_2$, SOC is expected to be weak and therefore interband pairing
should be considered.

\section{Classification of Superconducting States}

Now we perform a symmetry analysis of possible superconducting states.
LaNiC$_2$ crystallizes into a single phase of orthorhombic space group $Amm2$ (No.\ 38, $C_{2v}^{14}$). 
Possible superconducting states are derived from irreducible representations 
of the normal-state symmetry group.
If we consider that SOC is weak then there are two approaches: \\
i)  band splitting due to SOC is neglected, but the helicity basis is used to describe superconductivity, with interband pairing.  Then the normal-state
symmetry is $C_{2v}\times T\times U(1)$, where $T$ is time-reversal
and $U(1)$ is gauge (phase) symmetry.\cite{Quintanilla}  \\
ii) spin-orbit coupling is completely neglected and so the
symmetry group (for a pair of spins) is $SO(3)\times C_{2v}\times T\times U(1)$.
\cite{Hillier} \\
These approaches differ from the limit of strong SOC when interband pairing
can be neglected.\cite{Sergienko2004}

\begin{table}
\caption{\label{tab:D2h_SOC}Sample gap functions
for  singlet $\psi(\k)$ and triplet ${\bm d}(\k)$ pairing 
for the point group $C_{2v}$ when SOC is included \cite{Annett}.  The first
column lists the irreducible representations of $C_{2v}$, the second and third
columns list representative forms for the functions $\psi(\k)$ and
${\bf d}(\k)$, the fourth column 
is the symmetry of the superconducting state and the fifth 
column lists symmetry-required nodes.
$A$, $B$ and $C$ are arbitrary constants.}
\begin{tabular}{|c|c|c|c|c|}
\hline 
$C_{2v}$&
$\psi(\k)$&
${\bf d}(\k)$&
Symmetry & Nodes\tabularnewline
\hline
\hline 
$A_{1}$&
$1$&
$(Ak_x, Bk_y, Ck_z)$ &
$C_{2v}\times T$&  none \tabularnewline
\hline 
$A_{2}$&
$k_x k_y$&
$(Ak_y, Bk_x, Ck_x k_y k_z)$ &
$C_{2v}(C^z_{2}) \times T$ & point $[001]$  \tabularnewline
\hline 
$B_{1}$&
$k_x k_z$&
$(Ak_z, Bk_x k_y k_z, Ck_x)$ &
$C_{2v}(\sigma_{xz}) \times T$ & point $[010]$  \tabularnewline
\hline 
$B_{2}$&
$k_y k_z$&
$(Ak_x k_y k_z, Bk_z, Ck_x)$ &
$C_{2v}(\sigma_{yz})\times T$& point $[100]$  \tabularnewline
\hline
\end{tabular}
\end{table}
The first approach yields functions $\psi(\k)$ and ${\bf d}(\k)$  that are the same as 
strong SOC in a centrosymmetric superconductor, 
therefore the trial 
functions listed in Table \ref{tab:D2h_SOC} are the same as those 
for $D_{2h}$.\cite{Annett}
The only difference between the NCS and centrosymmetric cases is that
mixed parity states are allowed.
The magnitude of the gap of a mixed-parity centrosymmetric superconductor is 
$|\Delta_{\pm}(\k)|^2 = |\psi(\k)|^2 + |{\bm d}(\k)|^2 \pm |{\bm p}(\k) + {\bm q}(\k)|$, where 
${\bm p}(\k) = \psi(\k) {\bm d}^{*}(\k) + \psi^{*}(\k){\bm d}(\k)$
and ${\bm q}(\k)  = i {\bm d}(\k)\times {\bm d}^{*}(\k)$.\cite{Sergienko}  TRS is not broken for any of the phases involving  the 1D 
order parameters listed in Table \ref{tab:D2h_SOC}, but the gap function is
non-unitary because $\Delta \Delta^\dagger$ is not proportional to the identity matrix and the 
magnitude of the gap is $|\Delta_{\pm}(\k)| = |(|\psi(\k)| \pm |{\bm d}(\k)|)|$.
Symmetry-required nodes are those which occur in $\Delta_{\pm}(\k)$ for any values of the 
parameters $A$, $B$ and $C$; lines nodes may also be found for $\Delta_{-}(\k)$
depending on the choice of parameters.\cite{Sergienko}


\begin{table}
\caption{\label{tab:D2h_noSOC}Sample gap functions for  singlet  and triplet pairing for the point group $C_{2v}$ when there is no SOC.\cite{Annett}
The first column lists the irreducible representations of $SO(3) \times  C_{2v}$, the second column lists representative forms of $\psi(\k)$ and ${\bf d}(\k)$,
the third column lists the symmetries of the superconducting states 
and the fourth column lists symmetry-required nodes.}
\begin{tabular}{|c|c|c|c|}
\hline 
~~&
$\psi(\k)$&
Symmetry & Nodes \\
\hline
$^{1}A_{1}$&
$1$&
$SO(3)\times C_{2v} \times T$& none 
\\
\hline 
$^1A_{2}$&
$k_x k_y$&
$SO(3)\times C_{2v}(C^z_{2})\times T$&
lines\\ 
~~~&
~~~&
~~~&
$k_x,k_y=0$ \\
\hline 
$^1B_{1}$&
$k_x k_z$&
$SO(3)\times C_{2v}(\sigma_{xz})\times T$&
lines \\
~~~&
~~~&
~~~& 
$k_x,k_z=0$ \\
\hline
$^1B_{2}$& $k_y k_z$& $SO(3)\times C_{2v}(\sigma_{yz})\times T$& 
lines \\
~~~&
~~~&
~~~&
$k_y,k_z=0$\\
\hline 
\hline
~~& ${\bf d}(\k)$ & Symmetry& Nodes \tabularnewline
\hline
$^3A_{1}$&$ (0,0,1)k_x k_y k_z$ & $D_{\infty}(C_{\infty})\times C_{2v} \times T$& lines  \tabularnewline 
& $(1,i,0)k_x k_y k_z$ & $D_{\infty}(E)\times C_{2v} $ &  $k_x,k_y,k_z=0$ \tabularnewline
\hline
$^3A_{2}$& $(0,0,1)k_z$ & $D_{\infty}(C_{\infty})\times C_{2v}(C^z_{2}) \times T$& line \tabularnewline 
& $(1,i,0)k_z$ & $D_{\infty}(E)\times C_{2v}(C^z_{2})$ & $k_x = 0$\tabularnewline
\hline
$^3B_{1}$& $(0,0,1)k_y$ & $D_{\infty}(C_{\infty})\times C_{2}(\sigma_{xz}) \times T$ & line \tabularnewline
& $(1,i,0)k_y$& $D_{\infty}(E)\times C_{2v}(\sigma_{xz})$ &$ k_y = 0$ \tabularnewline
\hline 
$^3B_{2}$& $(0,0,1)k_x$ & $D_{\infty}(C_{\infty})\times C_{2v}(\sigma_{yz})  \times T$ &  line \\
& $(1,i,0)k_x$ & $D_{\infty}(E)\times C_{2v}(\sigma_{yz})$ & $k_x =0$ \tabularnewline
\hline 

\end{tabular}
\end{table}

The gap functions resulting from the second approach are tabulated in 
Refs.\ \onlinecite{Hillier,Quintanilla} and reproduced in
Table \ref{tab:D2h_noSOC}.   Table  \ref{tab:D2h_noSOC} also 
gives the symmetry\cite{Annett} and symmetry-required nodes of each possible phase.
In the singlet cases, $SO(3)$ spin symmetry is preserved.  However,
in the triplet cases two possibilities are realised.  In the first possibility
the triplet spin state is of the form $|\uparrow \downarrow\rangle 
+ |\downarrow \uparrow\rangle$.   The symmetry of this spin state is 
$D_{\infty}(C_{\infty})\times T$.  This possibility is shown in the 
upper rows of the triplet cases in 
Table \ref{tab:D2h_noSOC}. 
The other possibility is that the spin state is of the form
$|\uparrow \uparrow\rangle$, that is, the spin points in a definite direction.  
This state breaks TRS; its symmetry group is
$D_{\infty}(E)$.  Combined with the real-space part of the gap functions, 
we find twelve distinct superconducting phases.
In all cases except the trivial $^1A_1$ case the gap 
function has line nodes; in the non-unitary case these occur 
in $\Delta_{+}(\k)$, where $|\Delta_{\pm}(\k)| =\sqrt{|{\bm d}(\k)|^2 \pm |{\bm q}(\k)|}$
and ${\bm q}(\k) =  i {\bm d}(\k) \times {\bm d}^{*}(\k)$, while
the $\Delta_{-}(\k)$ is gapless.\cite{Hillier}

We now discuss those phases that have broken TRS, as observed in Ref.\
\onlinecite{Hillier}.
According to the symmetry classifications, 
the only phases that break TRS 
are found in the 
weak SOC classification scheme (Table \ref{tab:D2h_noSOC}) and are non-unitary. 
These always occur with line nodes 
{\em and}  gapless excitations.\cite{Hillier}
Gaplessness
should be reflected in power laws, but it seems unlikely that any of the early 
specific heat experiments\cite{Lee,Pecharsky} 
are compatible with gapless superconductivity.
Therefore, it is difficult to reconcile a state with broken TRS with these measurements.

There is second issue related to broken TRS in a $C_{2v}$ crystal that is difficult to resolve.
According to the results of symmetry classification, a TRS breaking state is possible only when SOC is completely neglected.   However, SOC exists in the normal state\cite{Hase}, and while its effects are smaller than that of 
CePt$_3$Si, it is large enough to justify a strong SOC approach.\cite{Hase} 
However, there are no broken TRS states in this approach.
Therefore, in order for TRS to be broken there must be a decoupling of the orbital and spin degrees of freedom in the superconducting phase.


To summarise,
beginning with the most general description of superconductivity in
non-centrosymmetric crystals,  
we have analysed possible superconducting phases for the point group $C_{2v}$, 
assuming weak spin-orbit coupling.  In order to account 
for TRS breaking, superconductivity in LaNiC$_2$ should be described 
by a non-unitary order parameter in the triplet channel with SOC neglected. 
Such a gap function has line nodes in its upper branch while its lower
branch is gapless. In order to confirm this phase, further investigations  
that can establish the existence and positions of line nodes and gapless superconductivity are required.

\end{document}